\documentclass[superscriptaddress,twocolumn,prb]{revtex4}%
\usepackage{graphicx}
\usepackage{dcolumn}
\usepackage{bm}
\usepackage{color}
\usepackage{ulem}
\usepackage{amsmath}
\usepackage{amsfonts}
\usepackage{amssymb}
\usepackage{tikz}

\begin{document}

\title{Spin-orbit Interaction driven Topological Features in a Quantum Ring}

\author{Shenglin Peng}
\affiliation{State Key Laboratory of Powder Metallurgy, and Powder Metallurgy
Research Institute, Central South University, Changsha, P. R. China 410083}
\affiliation{School of Physics and Electronics, Central South University,
Changsha, P. R. China 410083}

\author{Wenchen Luo }
\email{luo.wenchen@csu.edu.cn}
\affiliation{School of Physics and Electronics, Central South University,
Changsha, P. R. China 410083}

\author{Fangping Ouyang }
\email{ouyangfg@csu.edu.cn}
\affiliation{State Key Laboratory of Powder Metallurgy, and Powder Metallurgy
Research Institute, Central South University, Changsha, P. R. China 410083}
\affiliation{School of Physics and Electronics, Central South University,
Changsha, P. R. China 410083}
\affiliation{School of Physics and Technology, Xinjiang University, Urumqi,
P. R. China 830046}

\author{Tapash Chakraborty }
\email{Tapash.Chakraborty@umanitoba.ca}
\affiliation{Department of Physics and Astronomy, University of Manitoba,
Winnipeg, Canada R3T 2N2}

\date{\today }

\begin{abstract}
One-dimensional quantum rings with Rashba and Dresselhaus spin-orbit couplings
are studied analytically and are in perfect agreement with the numerical
results. The topological charge of the spin field defined by the winding
number along the ring is also studied analytically and numerically in the
presence of the spin-orbit interactions. We also demonstrate the cases where
the one-dimensional model is invalid for a relatively large radius. However,
the numerical results of the two-dimensional model always remain reliable.
Just as many physical properties of the quantum rings are influenced by the
Aharonov-Bohm effect, the topological charge is also found to vary
periodically due to the step-like change of the angular momentum with an
increase of the magnetic field. This is significantly different from the cases
of quantum dots. We also study how the current is induced by the magnetic
field and spin-orbit couplings, which is strong enough that it could to be
detected. The magnetic induction lines induced by the spin field and the
current are also analyzed which can be observed and could perhaps help
identifying the topological features of the spin fields in a quantum ring.
\end{abstract}

\maketitle

\section{Introduction}

In the study of topological properties of condensed matter states,
the crucially important role of the spin-orbit coupling (SOC) has
been recognized in recent years \cite{TP03,TP04,tokura,sky01,sky02}.
It is important in the transport properties of the quantum Hall
systems \cite{ezawa}, such as silicene/germanene \cite{si} and
bilayer graphene \cite{abergel,wang_2010,bilayerg} (where there
is pseudospin-orbit coupling). The nontrivial topological
bands in the momentum space explicitly lead to ballistic transport
at the edge of the topological insulator. Here we report on
the topology of the spin texture in real space and its influence
on the persistent current when the spin fields are topologically
nontrivial in a quantum ring. Additionally, the persistent current
and the spin fields induce effective magnetic fields which are
found to be strong enough and can be observed.

Spin textures in real space in both noninteracting and interacting
quantum dots \cite{maksym,QD_book,QD01,QD03,QD04,QD05,qd6}
with the SOCs have been studied recently \cite{luo1,luo2}. The winding
number was previously introduced to describe the topological charge $q$ of
the spin field, and the topology of the in-plane spin
texture can be tuned by the external electric and magnetic fields
\cite{luo1,luo2}. The detection of the topological charge in
low-dimensional systems is difficult but can be found in an indirect way,
i.e., by measuring the sign of the $z$ component of the spin
$\langle\sigma^{}_z\rangle$ in a large dot. Since the sign of
$\langle\sigma^{}_z\rangle$ can be inverse in weak magnetic fields
in the presence of the SOC the topological charge induced by
the SOC can then be determined \cite{luo1}. Experimental determination
of the topological winding number by polarized resonant X-ray
scattering process \cite{winding_nature} could also be a possibility.

The quantum ring \cite{ring1,Lucignano,chapter,maxwelldemon},
which is yet another two-dimensional (2D) nano-device, has
a similar Hamiltonian as that of the quantum dot. However, the quantum
ring has its own features that are essentially different from the dot,
since the geometric structure of the ring with different confining
potential in the Hamiltonian introduces the magnetic flux into the
system. Interestingly, we found that the one-dimensional (1D) model
is not appropriate in the analysis of spin textures in a ring of
large radius. It fails to explain the spin rotation in the radial
direction which can however be found numerically in the 2D model. It is
because the radial
degree of freedom is integrated out in the 1D model, but this degree
of freedom is important, especially when the radius of the ring is
large. This size effect is associated with the flip of $\langle
\sigma^{}_z \rangle$, that can be observed experimentally.

Spin textures with topological charges are expected to appear in a ring
due to the broken translational symmetry. Note that due to the geometry
differences, the spin textures in quantum rings are very different from
those in quantum dots. It can also be seen from the change of the $z$
component of the angular momentum, $\langle L^{}_z\rangle$. In the
single-electron case, $\langle L^{}_z \rangle$ in a quantum dot varies
from $0$ to $-1$ with the increase of the magnetic field $B$. Then the
topological charge may be changed from $q=\mathrm{sgn}(g)$ to $q=
-\mathrm{sgn}(g)$ where $g$ is the Land\'e factor of the material when
both of the SOCs are present \cite{luo2}. In contrast, $\langle L^{}_z
\rangle$ changes gradually to $-\infty$ in a ring. Hence, we
believe that the topological charge in a quantum ring will be changed more
than once since the wave function highly depends on the $\langle L^{}_z
\rangle$. In our analysis that follows, we shall see that $q$ is indeed
changed periodically with the increase of the magnetic field, as the
ground state is changed when the magnetic flux increases by a magnetic
flux quanta.

Considering the geometric structure of the ring, we then explore
the current induced by both the magnetic field and the SOCs. In
fact, the SOCs can be treated as an effective magnetic field, which
then induces a local current even at zero magnetic field. Our results
are comparable to the previous results where only one-dimensional ring
was considered \cite{sheng}. The two SOCs compete with each other to
determine the vorticity of the current field. We note that the current
flows locally but there is no net current, since the electron is confined
locally in the ring. The magnetic fields induced by the current and
the spin of the electron are also calculated in a semi-classical treatment.
Both the current and the induced magnetic fields may actually be detected,
and the measurements will be related to the topological features of
the spin fields.

The paper is organized as follows. In Section II, we introduce the
Hamiltonian of the quantum ring with Rashba and Dresselhaus SOCs, and define
the spin field and its winding number. The current field is also defined
there. In Section III, we simplify the Hamiltonian in the 1D limit. In a
finite magnetic field, the Hamiltonian with Zeeman coupling is analyzed
perturbatively. In Section IV, we study the spin textures of the electron
in a quantum ring. In fact, we analyze the spin textures in the one-dimensional
model and the results can be verified numerically in a two-dimensional ring.
We then evaluate the current in the first-order perturbation theory and
also numerically. It includes two parts: one that is induced by the
magnetic field while the other is induced by the SOCs. The magnetic fields
caused by the current and spin field are also calculated in Sec. III.C.
In Sec. IV, we explain the limitations of the 1D model. We show that
the spin rotates with the increase of the radius only when the ring is
two-dimensional. Finally, we close with summary and conclusion in Sect. V.

\section{Quantum ring model with the SOCs}

We assume that the ring is a perfect circle and the Hamiltonian of the
quantum ring of radius $r_0$ with SOCs is generally given by
\begin{align}
H & =\frac{\mathbf{P}^2}{2m^{\ast}}+\frac{m^{\ast}}2\omega^2\left(
r-r^{}_0\right)^2+\frac{\Delta}2\sigma_z^{}+H^{}_{SOC},\label{h}\\
H^{}_{SOC} & =g_1^{}\left(\sigma_x^{}P_y^{}-\sigma_y^{}P_x^{}\right)
+g_2^{}\left( \sigma_y^{}P_y^{}-\sigma_x^{}P_x^{}\right),
\end{align}
where $\omega$ describes the radial parabolic confinements. $\sigma^{}_i$ are
the Pauli matrices and the strengths of the Rashba and linear Dresselhas SOCs
are given by $g_1^{}$ and $g_2^{}$, respectively. $P_i^{}=p^{}_i+eA_i^{}$
is the canonical momentum plus the vector potential which is
set in the symmetric gauge $\mathbf{A}=\frac12B\left( -y,x,0\right) $ with
the magnetic field $B.$ The Zeeman energy is then $\Delta=g\mu_B^{}B$, where
$g$ is the Land\'e factor and $\mu^{}_B$ is the Bohr magneton. We then rewrite
the Hamiltonian in the form
\begin{align}
H  &  =H_0^{}+H_r^{}+H^{}_{SOC},\label{hamiltonian}\\
H^{}_0  &  =\frac{\mathbf{P}^2}{2m^{\ast}}+\frac{m^{\ast}}2\omega^2%
r^2+\frac{m^{\ast}}2\omega^2r_0^2+\frac{\Delta}2\sigma^{}_z,\\
H^{}_r  &  =m^{\ast}\omega^2r^{}_0r,
\end{align}
The unperturbed Hamiltonian $H^{}_0$ can be diagonalized by the Fock-Darwin
basis,
\begin{align}
\psi^{}_{n,l}\left(\bf{r}\right) & =\frac1{\ell}\sqrt{\frac{n!}
{\pi\left(n+\left\vert l\right\vert \right)!}}\exp\left(-il\theta\right)
\label{fd}\\
& \times\left( \frac{r}{\ell}\right)^{\left\vert l\right\vert }
L_n^{\left\vert l\right\vert}\left(\frac{r^2}{\ell^2}\right)
\exp\left(-\frac{r^2}{2\ell^2}\right), \nonumber \\
\psi^{}_{n,l,+}\left(\bf{r}\right) & =\psi^{}_{n,l}\left( \bf{r}\right)
\left(
\begin{array}
[c]{c}%
1\\
0
\end{array}
\right) ,\\
\psi^{}_{n,l,-}\left(\bf{r}\right) & =\psi^{}_{n,l}\left( \mathbf{r}\right)
\left(
\begin{array}
[c]{c}%
0\\
1
\end{array}
\right),
\end{align}
where $n$ is the Landau level index, $l$ is the angular momentum index,
$\ell=\sqrt{\hbar/(m^{\ast}\sqrt{\omega^2+\omega_c^2/4})}$ is the
natural length with the cyclotron frequency $\omega^{}_c=eB/m^{\ast}$, and $L$
is the Laguerre polynomial. The Fock-Darwin states in Eq.~(\ref{fd}) are the
basis in the exact diagonalization of the Hamiltonian.

\subsection{Spin fields}

By diagonalizing the Hamiltonian in the Fock-Darwin basis, we have the eigen
state
\[
\Psi\left( \mathbf{r}\right) =\sum^{}_{n,l,s}C^{}_{n,l,s}\psi^{}_{n,l,s}\left(
\mathbf{r}\right),
\]
where the coefficients $C^{}_{n,l,s}$ are obtained numerically. The spin fields
$\sigma^{}_{\mu}\left( \mathbf{r}\right) $ of such a state can be calculated by
\begin{equation}
\sigma^{}_{\mu}\left( \mathbf{r}\right) =\sum^{}_{n,l,s}\sum^{}_{n^{\prime},l^{\prime
},s^{\prime}}C_{n,l,s}^{\ast}C^{}_{n^{\prime},l^{\prime},s^{\prime}}\psi
_{n,l,s}^{\dag}\left( \mathbf{r}\right) \sigma^{}_{\mu}\psi^{}_{n^{\prime}%
,l^{\prime},s^{\prime}}\left( \mathbf{r}\right) ,\label{spin}%
\end{equation}
and the density is given by
\begin{equation}
n\left( \mathbf{r}\right) =\sum^{}_{n,l,s}\sum^{}_{n^{\prime},l^{\prime},s^{\prime}%
}C_{n,l,s}^{\ast}C^{}_{n^{\prime},l^{\prime},s^{\prime}}\psi_{n,l,s}^{\dag}\left(
\mathbf{r}\right) \psi^{}_{n^{\prime},l^{\prime},s^{\prime}}\left( \mathbf{r}%
\right) .\label{density}%
\end{equation}
The average value of the quantity is also given by $\left\langle
A\right\rangle =\int A\left( \mathbf{r}\right) d\mathbf{r}$. The in-plane
field can be described by the vector field $\mathbf{\sigma}\left(
\mathbf{r}\right) =\left( \sigma^{}_x\left( \mathbf{r}\right), \sigma^{}_y\left(
\mathbf{r}\right) \right).$

The winding number was introduced in Ref.~\cite{luo2}. In a quantum dot, the
contour of the integral could be any closed path around a singularity of the
$\mathbf{\sigma}$ field. However, in a quantum ring, it is natural to define
the contour the same as the ring. In general, we define the winding number
along the circle with the radius $r^{\prime}$
\begin{equation}
q\left(r^{\prime}\right) =\frac{1}{2\pi}\oint^{}_{r=r^{\prime}}\frac{\sigma
^{}_x\,\left( \mathbf{r}\right) d\sigma^{}_y\left( \mathbf{r}\right) -\sigma
^{}_y\,\left( \mathbf{r}\right) d\sigma^{}_x\left( \mathbf{r}\right) }%
{\sigma^{}_x\left( \mathbf{r}\right)^2+\sigma^{}_y\left( \mathbf{r}\right)
^2}.
\end{equation}
Since the density of the electron is localized around the ring for lowering
the potential energy and the topological feature is meaningful associated with
the electron density, we therefore consider $q(r^{}_0)$ as the natural topological
charge of the spin field in the ring.

We note here that the topological feature along the ring could be very
different from that at a distance from the center, that is, $q\left(r\right)$
depends on $r$. As a result, the topological features of the quantum ring are
significantly different comparing with those in a quantum dot. Moreover, the
spin texture lose its relevance when the charge density is small. To
eliminate the textures of the spin fields with low charge density, we
define a new parameter
\begin{equation}
q^{\prime}=\frac{1}{2\pi}{\oint\limits_0^{2\pi}}\frac{\sigma^{}_x\left(
\theta\right) d\sigma^{}_y\left(\theta\right)-\sigma^{}_y\left( \theta\right)
d\sigma^{}_x\left(\theta\right)}{\sigma^{}_x\left(\theta\right) ^{2}%
+\sigma^{}_y\left(\theta\right)^2},
\end{equation}
where the spin fields are integrated on $r$
\begin{eqnarray}
\sigma^{}_x\left(\theta\right) =\int\sigma^{}_x\left(r,\theta\right) dr, \
\sigma^{}_y\left(\theta\right) =\int \sigma^{}_y\left( r,\theta\right) dr.
\end{eqnarray}
This means that the textures with highest density of the spin fields are the
most important part in the topological feature.

Therefore, it is better to define the winding number as $q^{\prime}$ while
$q\left(r\right)$ can be used to study the change of the topological charge with
radius. Further, a spatial distribution of the winding number density can also
be defined to observe the location of the vortex core
\begin{equation}
q^{\prime\prime}\left(\mathbf{r}\right) =\frac{\partial}{\partial\theta}%
\frac{\sigma^{}_x\left(\mathbf{r}\right) d\sigma^{}_y\left( \mathbf{r}\right)
-\sigma^{}_y\left(\mathbf{r}\right) d\sigma^{}_x\left(\mathbf{r}\right)
}{\sigma^{}_x\left(\mathbf{r}\right)^2+\sigma^{}_y\left(\mathbf{r}\right)
^2}.
\end{equation}
Determining the exact zero points of $q^{\prime\prime}\left( \mathbf{r}\right)$
can lead us to the cores of the vortices of the spin field.

\subsection{Current and vorticity}

The current operators can be derived by $j^{}_{\mu}=-\dfrac{\delta H}{\delta A}$,
so that the current densities are given by
\begin{align}
j^{}_x\left(\mathbf{r}\right) & =\frac{e}{2m^{\ast}}\left[\psi^{\dag}%
P^{}_x\psi+\left(P^{}_x\psi\right)^{\dag}\psi\right] \nonumber\\
& -e\psi^{\dag}\left( g^{}_1\sigma^{}_y+g^{}_2\sigma^{}_x\right) \psi,\\
j^{}_y\left(\mathbf{r}\right) & =\frac{e}{2m^{\ast}}\left[ \psi^{\dag}%
P^{}_y\psi+\left(P^{}_y\psi\right)^{\dag}\psi\right] \nonumber\\
& +e\psi^{\dag}\left(g^{}_1\sigma^{}_x+g^{}_2\sigma^{}_y\right) \psi,
\end{align}
where $\psi$ is the wave function spinor, and the current field is defined by
$\mathbf{j}\left( \mathbf{r}\right)
=\left( j^{}_x\left(\mathbf{r}\right),j^{}_y\left(\mathbf{r}\right) \right)
$. To classify the current fields induced by different SOCs, we define the
vorticity of the spin field, $\bm{\omega}^{}_s=\bm{\nabla}\times\bm{\sigma}$
and the vorticity of the current, $\bm{\omega}^{}_j=\bm{\nabla}\times
\mathbf{j,}$ respectively.

The current has three terms
\begin{equation}
j^{}_{\mu}\left(\mathbf{r}\right)\equiv j^{}_{z \uparrow,\mu}\left(\mathbf{r}\right)
+j^{}_{z\downarrow, \mu}\left( \mathbf{r}\right)
+j^{}_{SOC, \mu}\left( \mathbf{r}\right),
\end{equation}
where
\begin{eqnarray}
j^{}_{z \uparrow, \mu}\left( \mathbf{r}\right)&=& \frac{e}{2m^{\ast}} \left[
\psi_{\uparrow}^* P^{}_{\mu}\psi^{}_{\uparrow} +\left( P^{}_{\mu}
\psi^{}_{\uparrow}\right)^* \psi^{}_{\uparrow}\right], \label{current+} \\
j^{}_{z \downarrow, \mu}\left( \mathbf{r}\right)&=& \frac{e}{2m^{\ast}} \left[
\psi_{\downarrow}^* P^{}_{\mu}\psi^{}_{\downarrow} +\left( P^{}_{\mu}
\psi^{}_{\downarrow} \right)^* \psi^{}_{\downarrow} \right], \label{current-}\\
j^{}_{SOC, x} &=& -e\psi^{\dag}\left( g^{}_1\sigma^{}_y+g^{}_2\sigma^{}_x\right)
\psi, \label{currentx}\\
j^{}_{SOC, y} &=& e\psi^{\dag}\left( g^{}_1\sigma^{}_x+g^{}_2\sigma^{}_y\right)
\psi. \label{currenty}
\end{eqnarray}
The wave function spinor is
$\psi=
\left(
 \begin{array}{cc}
   \psi^{}_{\uparrow} & \psi^{}_{\downarrow} \\
 \end{array}
\right)^T$, and $\uparrow, \downarrow$ are related to the eigenstates of the
spin operator $\sigma^{}_z$.

\section{Analytical solutions of the one-dimensional quantum ring}

We first consider the one-dimensional (1D) model where the ring is so narrow
that it approaches the limit of a 1D circle. This ideal model is simple and
allows us to show the physical pictures of the system clearly. In this 1D
model, the requirement of hermiticity of the Hamiltonian with the SOCs,
means we must take into account properly the confinement of the wave function
in the radial direction \cite{meijer}
\begin{equation}
H=\frac{\mathbf{P}^2}{2m^*}+\frac12 g \mu^{}_B B \sigma^{}_z+ H^{}_{SOC},
\end{equation}
where the kinetic terms are given by
\begin{eqnarray}
P^{}_x &=&-i\hbar \left(-\frac1{2r^{}_{0}}\cos \theta -\frac1{r^{}_0}\sin
\theta\frac{\partial}{\partial\theta}\right)-\frac{eB}2r^{}_0\sin
\theta, \notag\\
P^{}_y &=&-i\hbar \left(-\frac1{2r^{}_0}\sin \theta +\frac{1}{r^{}_0}\cos
\theta \frac{\partial }{\partial \theta }\right) +\frac{eB}{2}r^{}_0\cos
\theta, \notag
\end{eqnarray}
and the constant term $H^{}_r$ is dropped.

The Hamiltonian can be solved
perturbatively when the SOCs are weak. We divide the Hamiltonian
into two parts, the unperturbed 1D ring, and the perturbed part, the SOCs
terms. The eigen states of the unperturbed Hamiltonian $H^{}_0$ are
\begin{align}
\left\langle \mathbf{r}\right. \left\vert l,+\right\rangle  &  =\psi
_{+}^{\left( 0\right)}\left( \mathbf{r}\right) =\frac{\delta^{}_{r,r^{}_0}}%
{r^{}_0\sqrt{2\pi}}\left(
\begin{array}
[c]{c}%
e^{-il\theta}\\
0
\end{array}
\right) ,\\
\left\langle \mathbf{r}\right. \left\vert l,-\right\rangle  &  =\psi
_{-}^{\left(0\right) }\left(\mathbf{r}\right) =\frac{\delta^{}_{r,r^{}_0}}%
{r^{}_0\sqrt{2\pi}}\left(
\begin{array}
[c]{c}%
0\\
e^{-il\theta}%
\end{array}
\right),
\end{align}
with the eigen energies
\begin{equation}
E^{}_{l,\pm}=E^{}_0\left( N-l\right)^2\pm\frac{\Delta}2,\label{ringenergy}%
\end{equation}
where $N=\frac{eBr_0^2}{2\hbar}$ and $E^{}_0=\frac{\hbar^2}{2m^* r_0^2}$.
The ground state is chosen by the sign of the Zeeman coupling (or the Land\'e
factor). The energy spectrum indicates a typical Aharonov-Bohm (AB) effect,
and the integer $l$ representing the angular momentum increases gradually to
minimize the energy of the ground state when the magnetic field is increased.

The Hamiltonian of the SOCs is
\begin{equation}
H^{}_{SOC}=\left(
\begin{array}
[c]{cc}%
0 & P^{}_{+}g^{}_1-iP^{}_{-}g^{}_2\\
P^{}_{-}g^{}_1+iP^{}_{+}g^{}_2 & 0
\end{array}
\right),
\end{equation}
where $P^{}_{\pm}=P^{}_y\pm iP^{}_x$.
If $H^{}_{SOC}$ is a perturbation then we obtain the first-order perturbations
\begin{align}
\psi_{+}^{\left( 1\right) }  &  =G^{}_{1+}\left\vert l-1,-\right\rangle
-iG^{}_{2+}\left\vert l+1,-\right\rangle,\\
\psi_{-}^{\left( 1\right) }  &  =-G^{}_{1-}\left\vert l+1,+\right\rangle
-iG^{}_{2-}\left\vert l-1,+\right\rangle,
\end{align}
where
\begin{align*}
G^{}_{1\pm}  &  =\frac{\hbar}{r^{}_0}\frac{\left( \frac12 \pm (N-l)\right) g^{}_1%
}{E^{}_0\left( N-l\right)^{2}-E^{}_0\left( N-l\pm1\right)^2\pm\Delta},\\
G^{}_{2\pm}  &  =\frac{\hbar}{r^{}_0}\frac{\left( \frac12 \mp (N-l)\right) g^{}_2%
}{E^{}_0\left( N-l\right)^2-E^{}_0\left( N-l\mp1\right)^2\pm\Delta}.
\end{align*}
The wave functions with the first-order perturbation corrections are
$\psi^{}_{\pm}=\psi_{\pm}^{\left( 0\right) }+\psi_{\pm}^{\left( 1\right) }$. For
the ground state we need to choose $\psi^{}_{+}$ for the negative Land\'e $g$ factor
($\Delta<0$) and $\psi^{}_{-}$ for the positive Land\'e $g$ factor ($\Delta>0$),
respectively.

In a weak magnetic field, $N-l\rightarrow0$, and the in-plane spin fields are
given by
\begin{align}
\sigma_x^{\pm}\left( \mathbf{r}\right)   &  =\pm 2G^{}_{1\pm}\cos\theta \mp
2G^{}_{2\pm}\sin\theta, \label{sigmax} \\
\sigma_y^{\pm}\left( \mathbf{r}\right)   &  =\pm 2G^{}_{1\pm}\sin\theta \mp
2G^{}_{2\pm}\cos\theta. \label{sigmay}
\end{align}
Clearly, if only the Rashba SOC exists then $q=1$, since $\mathbf{\sigma}\left(
\mathbf{r}\right) =\pm 2G^{}_{1\pm}\delta^{}_{r,r^{}_0}\left(
\cos\theta,\sin\theta\right)$. On the other hand, if only the Dresselhaus SOC
is present, then $\mathbf{\sigma}\left( \mathbf{r}\right) = \mp 2G^{}_{2\pm}%
\delta^{}_{r,r^{}_0}\left( \sin\theta,\cos\theta\right)$, and $q=-1$. These
results are not surprising when compared with the spin textures in a quantum
dot.

We now discuss the cases when both of the two SOCs are simultaneously present.
In a quantum dot hydrogen (one confined electron), $\langle L^{}_z\rangle$
remains close to zero and the topological transition of the spin texture could
occur, but at most once. In a quantum dot helium (two confined electrons),
$\langle L^{}_z\rangle$ varies and the second topological transition could
happen. The significant difference in a quantum ring hydrogen is that the
ground state appears with increasing $\langle L^{}_z\rangle$ when the magnetic
field is increased. From the experience in the quantum dot cases,
we could suppose that the varied $\langle L^{}_z\rangle$ may cause topological
transitions more than once. We still employ the 1D model to consider the more
general case, $g^{}_1,g^{}_2\neq0$, however, in a finite magnetic field
$N-l\neq0$. In order to minimize the energy of the unperturbed Hamiltonian,
or choose the proper unperturbed ground state, the angular momentum is
required to be
\begin{equation*}
-\frac12<l-N<\frac12.
\end{equation*}
The transition of $\left\langle L^{}_z\right\rangle $ happens when
$l-N=\pm\frac12$. We consider the limit $l-N\rightarrow\left( -\frac12
\right)^{+},$ i.e., $B\rightarrow\frac{2\hbar}{er_0^2}\left[ l+\left(
\frac12\right)^{-}\right] $, then $G^{}_{2+},G^{}_{1-}\rightarrow0$ and
\begin{align}
\psi_{+}^{\left( 1\right) }  &  =G^{}_{1+}\left\vert l+1,-\right\rangle
=\frac{\hbar}{r^{}_0}\frac{g^{}_1}{-2E^{}_0+\Delta}\left\vert l+1,-\right
\rangle,\\
\psi_{-}^{\left( 1\right) } & =-iG^{}_{2-}\left\vert l-1,+\right\rangle
=i\frac{\hbar}{r^{}_0}\frac{g^{}_2}{2E^{}_0+\Delta}\left\vert l-1,+\right
\rangle.
\end{align}
The spin fields are
\begin{align}
\mathbf{\sigma}^{+}\left( \mathbf{r}\right) & =2G^{}_{1+}\left( \cos
\theta,\sin\theta\right), \ q^{+}=1\\
\mathbf{\sigma}^{-}\left( \mathbf{r}\right)   &  =2G^{}_{2-}\left( \sin
\theta,\cos\theta\right), \ q^{-}=-1.
\end{align}
Then we consider another limit, $l-N\rightarrow\left( \frac{1}{2}\right)
^{-},$ i.e., $B\rightarrow\frac{2\hbar}{er_0^2}\left[ l-\left( \frac12
\right)^{-}\right]$, which leads to $G^{}_{2-},G^{}_{1+}\rightarrow0$.
In the same manner, we have
\begin{align}
\mathbf{\sigma}^{+}\left( \mathbf{r}\right)   &  =-2G^{}_{2+}\left( \sin
\theta,\cos\theta\right), q^{+}=-1,\\
\mathbf{\sigma}^{-}\left( \mathbf{r}\right)   &  =-2G^{}_{1-}\left( \cos
\theta,\sin\theta\right), q^{-}=1.
\end{align}
Therefore, it is clear that in the region $\frac{2\hbar}{er_0^2}\left(
l-\frac12\right) <B<\frac{2\hbar}{er_0^2}\left( l+\frac12\right)$ the
topological charge must be transformed from $-1$ to $1$ (or from $1$ to $-1$).

There is another topological transition at $B=\frac{2\hbar}{er_0^2}\left(l+
\frac12\right)$, since the topological charge is $q^{\pm}=\pm1$ for $B=
\frac{2\hbar}{er_0^2}\left[ l+\left( \frac12\right)^{-}\right]$ while
$q^{\pm}=\mp1$ for $B=\frac{2\hbar}{er_0^2}\left[l+\left(\frac12\right)^{+}
\right] =\frac{2\hbar}{er_0^2}\left[\left(l+1\right)-\left(\frac12\right)^{-}
\right].$ 

In conclusion, with the increase of the magnetic field, the sign of the
topological charge $q$ will be reversed in the region $\frac{2\hbar}{e
r_0^2}\left(l-\frac12\right)<B< \frac{2\hbar}{er_0^2}\left( l+\frac12\right)$. 
The sign of $q$ will be then inversed again at $B=\frac{2\hbar
}{er_0^2}\left( l+\frac12\right)$. In the next step, $q$ changes its
sign in the region of the magnetic field $\frac{2\hbar}{er_0^2}\left(
l+1-\frac12\right)<B< \frac{2\hbar}{er_0^2}\left( l+1+\frac12\right)$,
and then at $B=\frac{2\hbar}{er_0^2}\left(l+1+\frac12\right)$. When the
magnetic field continues to increase, $l$ also increases, and the topological
charge varies periodically. We will see in the next section that the numerical
results of a two-dimensional ring also support this conclusion.

\section{The spin textures and the induced current}

\subsection{Numerical results of the spin textures of a two-dimensional
quantum ring}

In order to generalize and verify the topological properties obtained in the
ideal one-dimensional model, we solve the Hamiltonian of a two-dimensional
quantum ring numerically. We consider the two-dimensional InAs quantum rings
with the parameters of the material: the effective mass $m^{\ast}=0.042
m^{}_e,$ the Land\'e factor $g=-14.6$, $r^{}_0$=15 nm, and $\hbar\omega$=40
meV. We use the exact diagonalization scheme to obtain the lowest states of
the system, in the basis of Fock-Darwin states $\left\vert n,l,s\right\rangle$
where $s=\pm$ is the spin index. We consider 460 Fock-Darwin states to
guarantee the accuracy of the calculations. In what follows, we use these
settings unless otherwise stated.

\begin{figure}[tbh]
\centering\includegraphics[width=3.23in,height=2.92in]{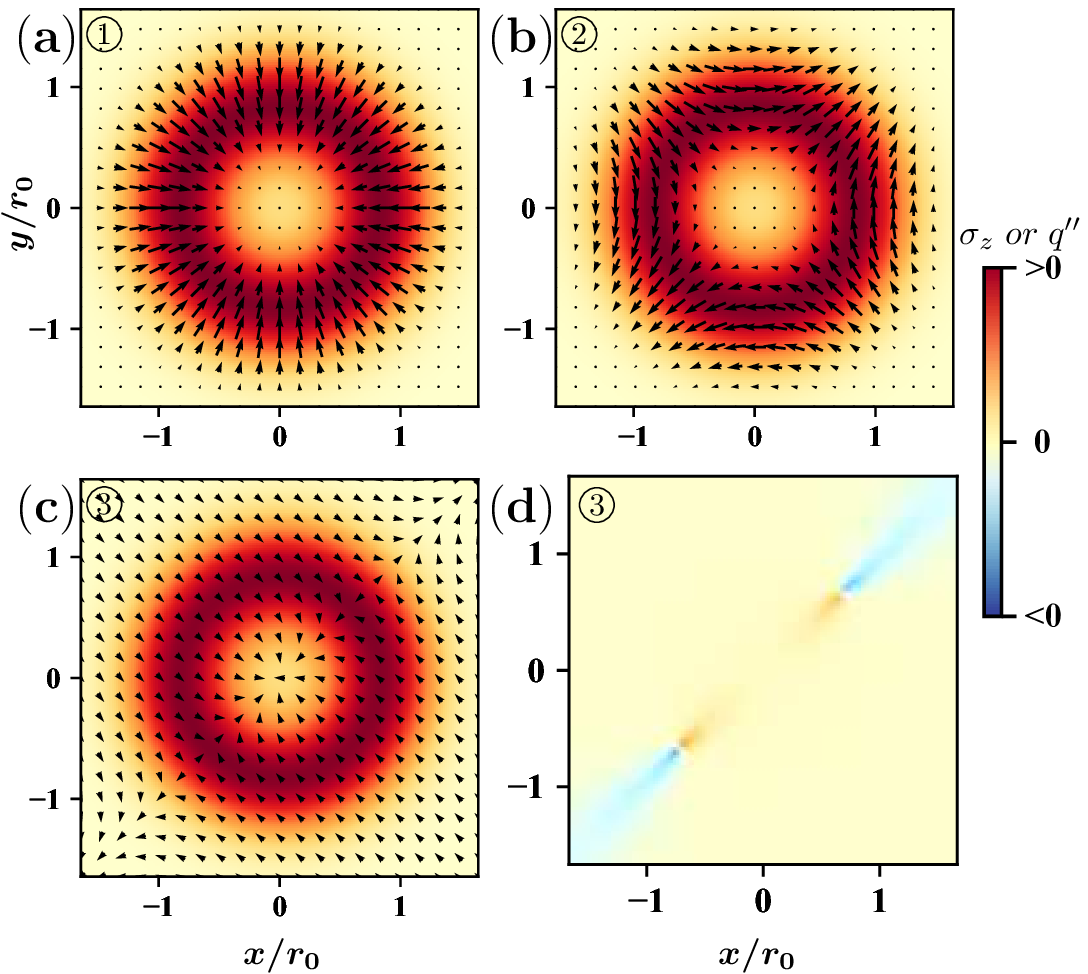}\caption{(color
online). Spin textures of the ground state of InAs rings, a) with Rashba SOC
only, $\hbar g^{}_1$ =20.0 nm$\cdot$ meV, b) with Dresselhauss SOC only, $\hbar
g^{}_2$ = 20.0 nm$\cdot$ meV, c) and with both SOCs, $\hbar g^{}_1=7.6$ nm$
\cdot$meV, $\hbar g^{}_2=12.4$ nm$\cdot$ meV at $B=2.8$T, and d) the spatial
distribution of the topological charge $q^{\prime\prime}\left(\mathbf{r}
\right)$ of Fig.~\ref{fig1} c). Hereafter, the arrows in the spin texture
figures represent the in-plane spin field $\mathbf{\sigma}$ and the background
colors represent the $\sigma^{}_z\left( \mathbf{r}\right)$, unless otherwise
stated. The arrows in Fig.~\ref{fig1} c) are normalized $\mathbf{\sigma}$
field, for $\mathbf{\sigma }\left( \mathbf{r}\right) $ is very small near
$\theta=0.25\pi$ and can not be shown clearly. The encircled numbers are
explained in the text.}
\label{fig1}
\end{figure}

Some typical spin textures in weak magnetic fields, which are similar to those
in quantum dots, are indicated in Fig.~\ref{fig1}. In Fig.~\ref{fig1}(a) we show
a vortex with $q = 1$, since only the Rashba SOC is present. In Fig.~\ref{fig1}
(b) a vortex with $q = -1$ is indicated when only the Dresselhaus SOCs is
present. The spin textures are also the same as for the exact solution and the
perturbation results discussed in the last section. Fig.~\ref{fig1}(c) shows
a vortex with $q=1$ if $r<r^{}_0$, and $q=-1$ if $r>r^{}_0$, when $\hbar g^{}_1
= 7.6$nm$\cdot$ meV, $\hbar g^{}_2 = 12.4$nm$\cdot$meV at $B=2.8$T. In this
example, we can see clearly that the topological charge depends on the path
of the integral. It is because, there are more than one vortices in
the plane. The $q''(\mathbf{r})$ is displayed in Fig.~\ref{fig1}(d) and its
zero points $(-0.8,-0.8), (0.8,0.8)$ and $(0,0)$ correspond to three vortex
cores. It is only natural then to calculate the winding number along the routes
exactly defined along the ring. Hence, we consider  the topological charge to
be given by $q'$, which is equivalent to $q(r^{}_0)$, in the case when the
ring is narrow.

\begin{figure}[tbh]
\centering\includegraphics[width=3.19in,height=3.33in]{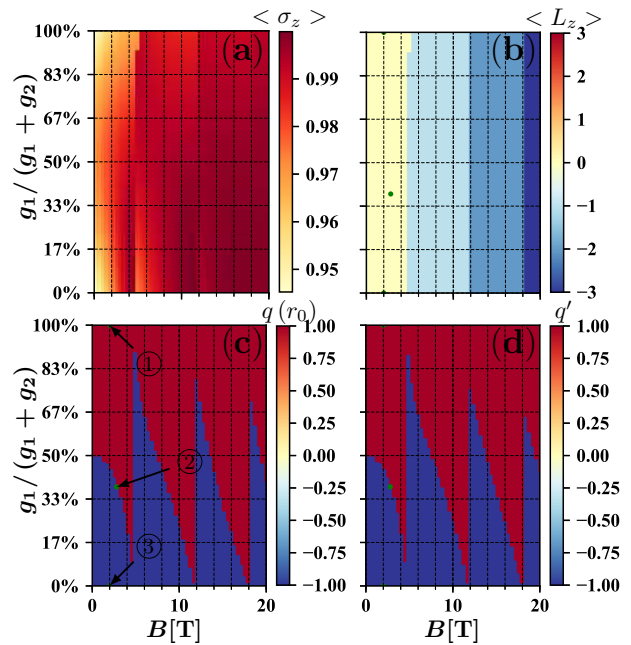}\caption{(Color
online) The (a) $\left\langle \sigma^{}_z\right\rangle$, (b) $\left\langle
L^{}_z\right\rangle$, (c) $q\left( r^{}_0\right)$ and (d) $q^{\prime}$ of the
ground state of InAs rings in $B\in${[}0, 20{]}T, $\hbar g^{}_1\in${[}0,
20{]} nm$\cdot$ meV and fixed $\hbar g^{}_1+\hbar g^{}_2$ = 20 nm$\cdot$ meV.}
\label{fig2}
\end{figure}

If the electron density is located around $r=r^{}_0$, the winding number along
the path $r=r^{}_0$ is consistent with the overall topological properties. As
shown in Fig.~\ref{fig2}, there is no major difference between $q\left(r^{}_0
\right)$ and $q^{\prime}$. The spin textures of the three points marked on
Fig.~\ref{fig2} (c) have been given in Fig.~\ref{fig1} (a)-(c).

Just as we found in the 1D ring, the topological charge can be varied
periodically with the increase of the magnetic field. Here, the numerical
results of the 2D rings also show that the topological features are tunable by
adjusting the magnetic field or the SOCs. As expected from the analytical
results, the topological charge of the ground state (Fig.~\ref{fig2} (d)),
following the change of $l$ which is the quantum number of the angular
momentum $\left\langle L^{}_z\right\rangle $ (Fig.~\ref{fig2} (b)), is also
periodically changed with the magnetic field. This is interesting since the
sign change in quantum dots happens at most once in quantum dot hydrogen or
helium \cite{luo2}. It is because $\left\langle L^{}_z\right\rangle $ is
varied when the magnetic flux of the ring is increased. Moreover, $\left
\langle \sigma^{}_z\right\rangle $ (Fig.~\ref{fig2}(a)) also changes with the
change of $\left\langle L^{}_z \right\rangle $, which can
be detected by NMR \cite{sean}. It is worth noting that the sign change of
$q^{\prime}$ does not occur exactly at $l-N=\pm\frac12$. This is because
the charge density in a 2D ring is not all concentrated at $r=r^{}_0$,
which is slightly biased from the strict 1D results. 

\subsection{Current induced by the external magnetic field and the SOCs}

\begin{figure*}[ptb]%
\centering
\includegraphics[
scale=0.9]%
{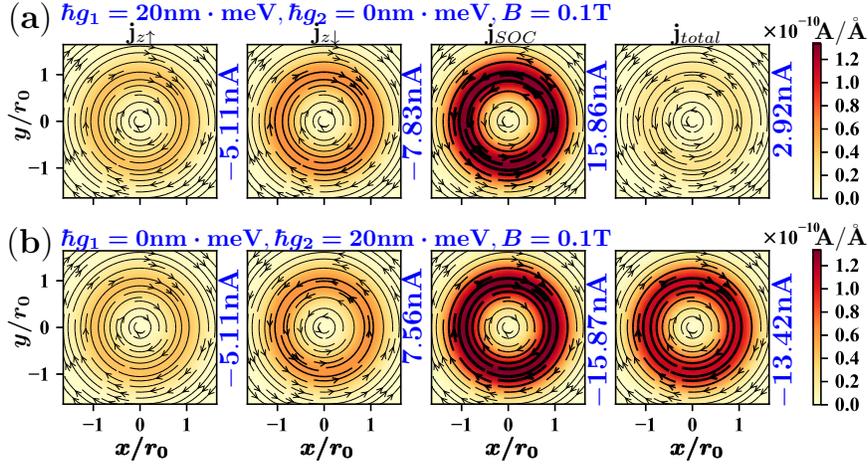}
\caption{(color online). Persistent current density of the ground states of
InAs rings when only one SOC exists: (a) for the Rashba SOC and (b) for the
Dresselhaus SOC. The parameters of the ring are indicated in the figures.
From left to right, $\mathbf{j}^{}_{z\uparrow}\left(\mathbf{r}\right)$,
$\mathbf{j}^{}_{z\downarrow}\left(  \mathbf{r}\right)$, $\mathbf{j}^{}_{SOC}
\left(\mathbf{r}\right)$ and the total current density $\mathbf{j}^{}_{total}
\left(\mathbf{r}\right)$. The currents $I^{}_{total}$ pass through the line
$x>$0, $y$=0 (clockwise is positive) are shown on the right side of the
figures. The background colors represent the absolute value of the current
density.}
\label{fig3}
\end{figure*}

\begin{figure*}[ptb]%
\centering
\includegraphics[
scale=0.9]%
{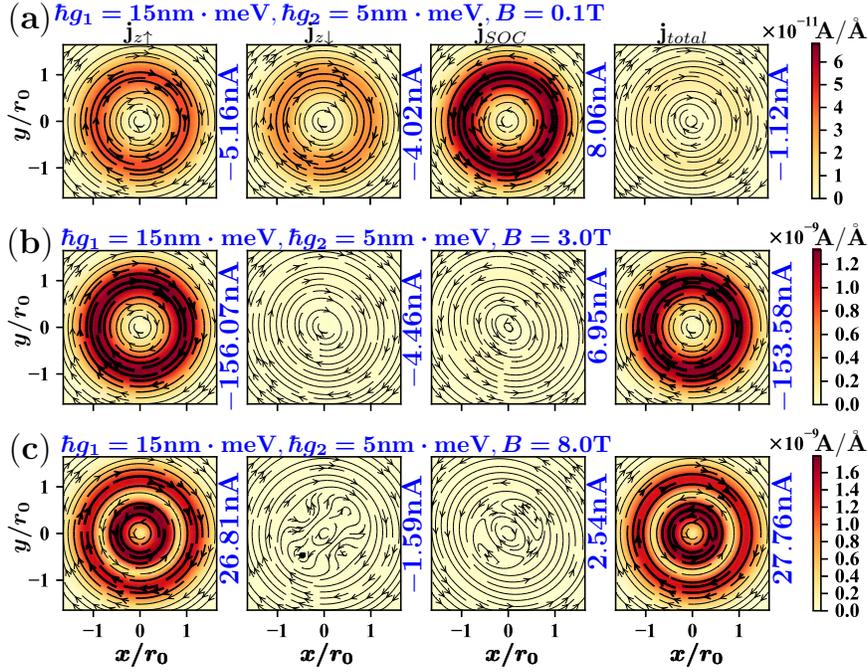}
\caption{(color online). Current density of the ground states of
InAs rings when both of the SOCs are present. (a)-(c) The parameters of the
ring are indicated in the figures. The order of the current fields is the same
as in Fig.~\ref{fig3}. $I^{}_{total}$ pass through the line $x>$0, $y$=0
(clockwise is positive) and is also shown. The colors represent the absolute
value of the current density.}
\label{fig4}%
\end{figure*}

In a quantum dot the Hamiltonian is
\begin{equation}
H^{}_{dot}=H_{0}^{}+H^{}_{SOC},
\end{equation}
(Eq.~(\ref{hamiltonian}) but without $H^{}_{r}$). The net current
$\left\langle j^{}_x\right\rangle =\left\langle j^{}_y\right\rangle =0$ since
the electrons are trapped locally in the dot. The conservation of the current
also requires that the net current vanishes. However, the local current density
could be nonzero. We perform the perturbation calculation where $H^{}_{SOC}$
is the perturbation, and are able to obtain the current density. We
suppose the Land\'e $g$ factor of the system $g<0$ (such as in InAs).
If there is no Dresselhaus SOC, then $g^{}_1\neq0,g^{}_2=0$, and we obtain
$\mathbf{j}^{R} = \frac12 e g^{}_1 \left( -\sigma_y^R\left( \mathbf{r}\right),
\sigma_x^R\left( \mathbf{r}\right) \right).$ This means that the current field
is perpendicular to the spin field $\mathbf{j}^R \cdot \bm{\sigma}=0$. The
local current does not vanish. The net current is zero since $\left\langle
\sigma^{}_{x,y}\right\rangle=0$.

For the Dresselhaus SOC, $g^{}_1=0,g^{}_2\neq0$, and we have
$\mathbf{j}^D = \frac12 e g^{}_2 \left( -\sigma_x^D\left( \mathbf{r}\right),
\sigma_y^D\left( \mathbf{r}\right) \right)$. The current is not perpendicular
to the spin field in this case. We already know that the spin field has
topological charge $q=1$ for the dot with Rashba SOC, and $q=-1$ for a dot
with Dresselhaus SOC. The vorticity of the spin field is zero,
$\bm{\omega}^{}_s=\bm{\nabla} \times \bm{\sigma}=0$, if only one SOC is present.
However, the vorticity of the current of the Rashba dot $\bm{\omega}_c^R$
(or $\bm{\omega}_c^D$ of the Dresselhaus dot) is nonzero. The vorticities of
the two different spin-orbit coupled dots are just the opposite in weak magnetic
fields $B \rightarrow 0$, $\bm{\omega}_c^D/g_2^2=-\bm{\omega}_c^R/g_1^2$.

In a quantum ring the results are essentially equivalent to the case of the
dots in weak magnetic field. We again employ the 1D model and find that ($g<0$)
\begin{eqnarray}
j^{s}_x &=&-\frac{e}{m^* r_0^2} \xi^{}_{l,s}\sin\theta + s\frac{e}
{\pi r_0^2} \left[ g^{}_1\left( G^{}_{1 s}\sin \theta -G^{}_{2 s}
\cos \theta \right) \right.  \notag \\
&&+ \left. g^{}_2\left( G^{}_{1 s}\cos \theta -G^{}_{2 s}\sin
\theta \right) \right], \label{jx} \\
j^{s}_y &=&\frac{e}{m^* r_0^2} \xi ^{}_{l,s}\cos \theta - s\frac{e}{\pi
r_0^2}\left[ g^{}_1\left( G^{}_{1s}\cos \theta -G^{}_{2s}\sin \theta \right)
\right.  \notag \\
&&+ \left. g^{}_2\left( G^{}_{1s}\sin \theta -G^{}_{2s}\cos
\theta \right) \right], \label{jy}
\end{eqnarray}
where
\begin{eqnarray}
\xi^{}_{l,s} &=&\left(l-s\right)
G_{1s}^2+\left(l+s\right) G_{2s}^2-2 l G^{}_{1s}G^{}_{2s}\sin 2\theta \notag \\
&& + l -\frac12\frac{r_0^2}{\ell_B ^2}-\frac12\frac{r_{0}^2}{\ell_B^2}\left(
G_{2s}^2+G_{1s}^2-2G^{}_{1s}G^{}_{2s}\sin 2\theta \right) \notag
\end{eqnarray}
with $s=\pm $ corresponding to the case $g<0, g>0$, respectively. The magnetic
length is $\ell^{}_B=\sqrt{\hbar/(eB)}$.

For simplicity, we consider a ring with the Rashba spin-orbit interaction
only ($g^{}_2=0$) in a small magnetic field. Then $\ell^{}_B \rightarrow
\infty$ and $l=0$, so that the current is dominated by the SOC field
\begin{eqnarray}
\mathbf{j}^R &=& \frac{e}{2\pi} \frac{m^* g_1^2}{\hbar r^{}_0} (-\sin \theta,
\cos \theta).
\end{eqnarray}
The ring with Dresselhaus SOC only ($g^{}_1=0$) in a small magnetic field gives
the current field
\begin{eqnarray}
\mathbf{j}^D &=& \frac{e}{2\pi} \frac{m^* g_2^2 }{\hbar r^{}_0} (\sin \theta,
-\cos \theta).
\end{eqnarray}
It is obvious that these two currents are, as in the quantum dots, just the
opposite, and the vorticities of the two current fields are also opposite,
$\bm{\omega_c}^R/g_1^2=-\bm{\omega_c}^D/g_2^2$.

The corresponding numerical results of 2D rings with the Rashba SOC and with
the Dresselhaus SOC in magnetic fields are displayed in Fig.~\ref{fig3}. If
$B\rightarrow 0$, the numerical results perfectly agree with the results of
the 1D model shown above, $\mathbf{j}^{}_{z\uparrow} \approx0$,
$\mathbf{j}^{}_{SOC}$=$-2\mathbf{j}^{}_{z\downarrow}$, and
$\mathbf{j}^{}_{total}$=$0.5\mathbf{j} ^{}_{SOC}$, and the vorticity of
$\mathbf{j}^{}_{SOC}$ for different SOC is opposite. When the magnetic field
is very weak, the direction of the total current is determined by the
$\mathbf{j}^{}_{SOC}$ term. When the magnetic field is increased to B=0.1T
as shown in Figs.~\ref{fig3} (a) and (b), $\mathbf{j}^{}_{z\uparrow}$ is no
longer negligible. Note that the directions of $\mathbf{j}^{}_{z\uparrow}$
and $\mathbf{j}^{}_{z\downarrow}$ are the same and does not depend on the
type of SOC, which makes the total currents different in a finite magnetic
field (Figs.~\ref{fig3}(a) and (b)).

The sign of $\mathbf{j}^{}_{total}$ is easy to reverse when introducing the
Dresselhaus SOC (Fig.~\ref{fig4}(a)), since the direction of
$\mathbf{j}^D_{SOC}$ is opposite to the direction of $\mathbf{j}^R_{SOC}$.
When the magnetic field continues to increase (for instance up to $B=3$T),
$\mathbf{j}^{}_{z\uparrow}$ will play a major role (Fig.~\ref{fig4}(b)). We
also indicate the current field at $B=8$T in Fig. \ref{fig4}(c), where the 
effect of the SOCs is
weakened by the magnetic field, since the spin is more polarized in a stronger
magnetic field. The current induced by the SOCs is also weakened, but the
total current becomes complex for more than one circles.

\begin{figure}[ptb]%
\centering
\includegraphics[
height=2.65in,
width=3.1236in
]
{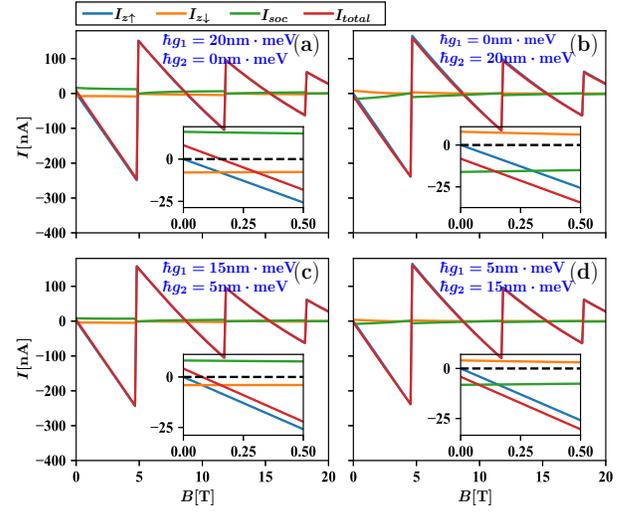}%
\caption{(Color online) In-plane persistent current of the ground state of
InAs rings a)-d). The parameters of the SOCs are indicated in the figures.}%
\label{fig5}%
\end{figure}

It is natural to define the current
\begin{equation}
I^{}_{\alpha}=\int\nolimits_{0}^{\infty}\mathbf{j}^{}_{\alpha}\left( x,y=0
\right) dx.
\end{equation}
The periodic oscillations of $I^{}_{total}$ induced by the AB effect can be
found in Fig.~\ref{fig5}, which is compatible with the previous works
\cite{sheng,janine}. Interestingly, $I^{}_{SOC}=-2I^{}_{z\downarrow}$, and
if $ g^{}_1+ g^{}_2$ is fixed, $I^{}_{z\uparrow}$ for any combination of
$g^{}_1$ and $g^{}_2$ are almost equal. At the points $l-N\approx0$,
$\ I^{}_{z\uparrow} \rightarrow 0.$ Then $I^{}_{SOC}$ plays an important
role, which makes it possible to distinguish the type of the SOCs by detecting
the direction of the current due to the fact that the direction of
$I^{}_{SOC}$ for Rashba and Dresselhaus SOCs are the opposite.

\subsection{Magnetic field induced by the current and spins}
\label{btoj}

The nonzero local current density can induce a magnetic field $\mathbf{B}^c$.
The steady current in the 2D ring flows along a couple of concentric circles
and induces the magnetic field in the three-dimensional (3D) space. In order to
calculate the induced magnetic field, we cover the plane of the ring by a
square lattice with the lattice constant $b$. According to the Biot-Savart
law, on the $i$-th site of the lattice the current induces the magnetic field
distributed in the space, which is given by
\begin{equation} \label{bslaw}
\mathbf{B}^c_i (\mathbf{r})=\frac{\mu^{}_0}{4\pi} \dfrac{b^2 \mathbf{j}
(\mathbf{r'}^{}_i) \times (\mathbf{r-r'}^{}_i)}{| \mathbf{r-r'}^{}_i|^3},
\end{equation}
where $\mu^{}_0$ is the vacuum permeability, $\mathbf{r}$ is a point in the
3D space, and $\mathbf{r'}^{}_i$ is the position of the $i-$th
site of the lattice. The magnetic field $\mathbf{B}^c$ induced by the
current is consequently calculated by the summation of the magnetic fields
induced by all the sites, $\mathbf{B}^c (\mathbf{r})= \sum^{}_i \mathbf{B}^c_i
(\mathbf{r}) $, once we obtain the full current field.
Meanwhile, the spin field also induces a magnetic field
$\mathbf{B}^s$ in the semiclassical treatment. However, in this case, we
need to introduce the $z$ axis into the system, since the spin field
is equivalent to the magnetic moment which is induced by the closed
current flowing on the plane perpendicular to the 3D spin field.

\begin{figure}[ptb]%
\centering
\includegraphics[
scale=0.7
]
{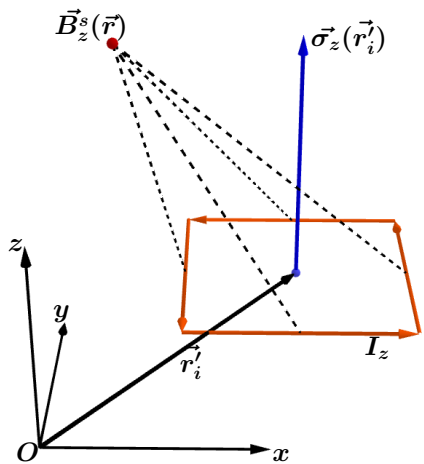}%
\caption{(Color online) The schematic diagram that how the spin field induces
an effective magnetic field semi-classically. The $z$ component of the spin
field $\sigma^{}_z(\mathbf{r'}^{}_i)$ at site $\mathbf{r'}^{}_i$ is equivalent
to the current square (orange) with side length $b$, and the currents induce
the magnetic field $\mathbf{B}^s_z(\mathbf{r})$ in the real space calculated by
the Biot-Savart law.}
\label{fig6}
\end{figure}

For each component of the spin field $\sigma^{}_{\nu}(\bm{r'}^{}_i)$ at the
site $i$, it is equivalent to a constant current along the square coil (with
the side length $b \ll \ell$) located in the plane perpendicular to the axis of
$\nu$. The constant current $I^{}_\nu$ induces a magnetic moment $\mu^{}_{
I^{}_{\nu}}$ at $\bm{r'}^{}_i$, which needs to be equal to the $\nu$ component
spin magnetic moment at $\bm{r'}^{}_i$,
\begin{equation}
\mu^{}_{I^{}_\nu} (\mathbf{r'}^{}_i) = I^{}_\nu b^2= \mu^{}_B \int^{}_{b
\times b} \psi^\dag \sigma^{}_{\nu} \psi dS ,
\end{equation}
where the two-dimensional integral covers the area of the $b\times b$ square.
When $b$ is very small we can approximately find the current
$I^{}_\nu = \mu^{}_B  \psi^\dag \sigma^{}_{\nu} \psi.$
Once the current $I^{}_\nu$ is found, we can use the Biot-Savart law to
calculate the magnetic field induced by the current along the edge of this
$b\times b$ square. The shcematic picture of this process is shown in
Fig.~\ref{fig6}, where we take the spin field $\sigma^{}_z (\mathbf{r}'_i)$
as an example. Then we sum over all the squares covering the area of the spin
field to obtain the magnetic field distribution $\mathbf{B}^s_\nu
(\mathbf{r})$. For instance, the magnetic field induced by the $z$ component
of the spin magnetic moment is given by
\begin{eqnarray}
\mathbf{B}_z^s\left( \mathbf{r}\right) &=&\frac{\mu ^{}_0}{4\pi }%
I^{}_zb\sum^{}_i\left[ \frac{\widehat{x}\times \left( \mathbf{r}-\mathbf{r}%
_i^{\prime }+\frac{b\widehat{y}}2\right) }{\left\vert \mathbf{r}-\mathbf{%
r}_i^{\prime }+\frac{b}2\widehat{y}\right\vert^3}\right.   \notag \\
&&+\frac{\widehat{y}\times \left( \mathbf{r}-\mathbf{r}_i^{\prime }-\frac{b%
\widehat{x}}2\right) }{\left\vert \mathbf{r}-\mathbf{r}_i^{\prime }-%
\frac{b}2\widehat{x}\right\vert^3}+\frac{-\widehat{x}\times \left(
\mathbf{r}-\mathbf{r}_i^{\prime }-\frac{b}2\widehat{y}\right) }{%
\left\vert \mathbf{r}-\mathbf{r}_i^{\prime }-\frac{b}2\widehat{y}%
\right\vert^3} \notag \\
&&\left. +\frac{-\widehat{y}\times \left( \mathbf{r}-\mathbf{r}_i^{\prime
}+\frac{b}2\widehat{x}\right) }{\left\vert \mathbf{r}-\mathbf{r}%
_i^{\prime }+\frac{b}2\widehat{x}\right\vert ^{3}}\right]
\end{eqnarray}
In a similar way we can calculate the other two components of the induced
magnetic field. Note that for $\sigma^{}_{x,y}$, we need to use the squares in
the $y-z$ and $x-z$ planes, respectively. Since $b \ll \ell$ and we do not
expand our rings too much in the $z$ direction, it is acceptable that the
ring is supposed to have a thickness $b$ which is much smaller than the radius
of the ring. The total magnetic field induced by the spin is thus given by
$\mathbf{B}^s (\mathbf{r})=\Sigma^{}_\nu \mathbf{B}^s_\nu (\mathbf{r})$.

\begin{figure*}[ptb]%
\centering
\includegraphics[
scale=0.85]
{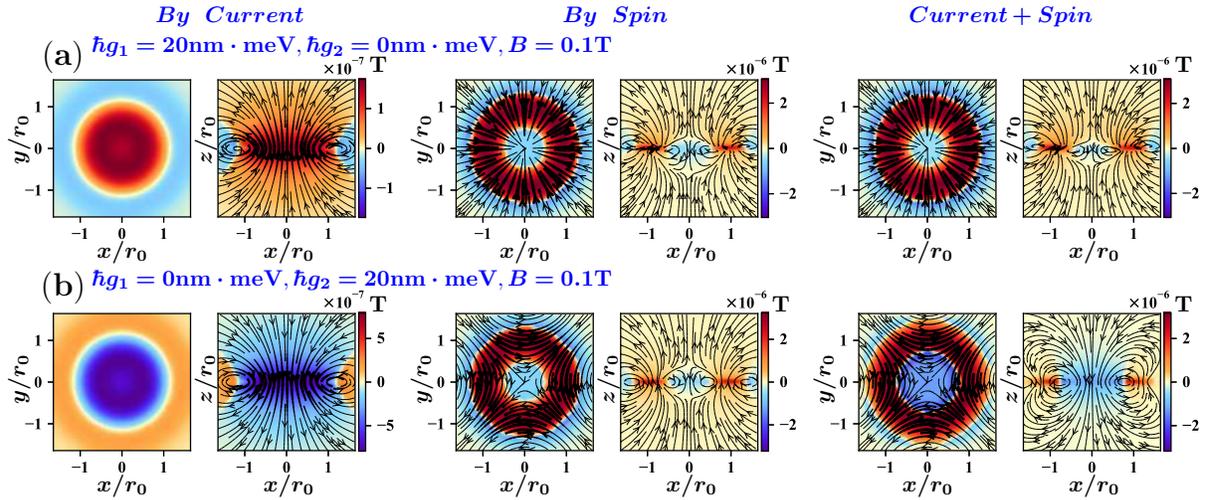}%
\caption{(color online). Magnetic fields induced by the current and the spins
with only one SOC. The parameters of the systems in (a) and (b) are the same
as those in Figs.~\ref{fig3}(a) and (b), respectively. The colors represent
the $z$ component of the induced magnetic field,
$\left\vert \mathbf{B}^{c,s}\right\vert \cdot \text{sgn} \left( B^{c,s}_z\right)$.
The lines in the first, third and fifth columns are the induced magnetic
induction lines on the $xOy$ plane.
The second, fourth and the sixth columns indicate the induced magnetic
induction lines on the $xOz$ plane.
}
\label{fig7}%
\end{figure*}

\begin{figure*}[ptb]%
\centering
\includegraphics[
scale=0.85]
{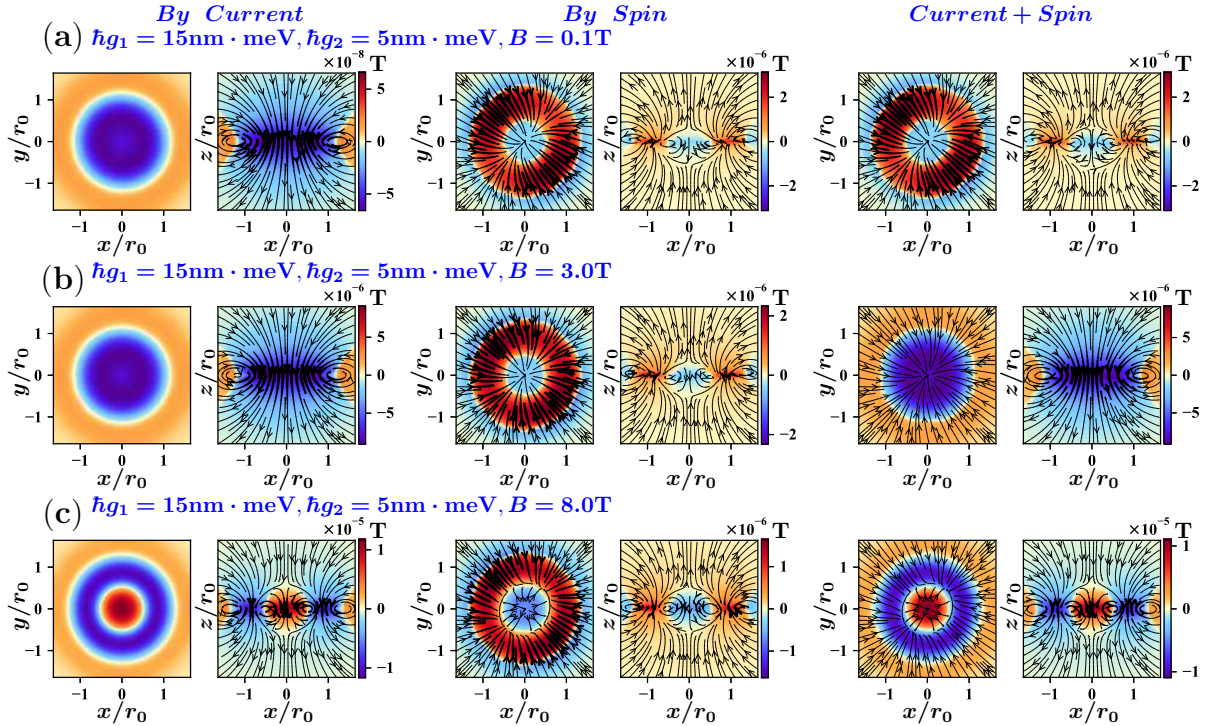}%
\caption{(color online). Magnetic fields induced by the current and the spins,
when both of the SOCs are present. The parameters of the systems in (a) to (c)
are the same as those in Fig.~\ref{fig4} (a) to (c) respectively. The arrows
represent the induced magnetic induction lines.
Colors represent $\left\vert \mathbf{B}^{c,s}\right\vert \cdot \text{sgn}\left(
B^{c,s}_z\right)$.
}
\label{fig8}%
\end{figure*}

When $B\rightarrow 0$, the direction of the current is opposite for different
SOCs, so that the current induced magnetic fields are opposite.
But $B^s$ is almost the same along the $z$ direction. Hence,
$\mathbf{B}^c +\mathbf{B}^s$ is distinguishable for different type of the
SOC (Fig. \ref{fig7}(a) for the Rashba SOC and Fig. \ref{fig7}(b) for the
Dresselhuas SOC), which can be utilized in indirectly determining topological
charge of the spin textures, if the induced magnetic field can be detected.
Since the direct measurement of the topology of the spin field is very
difficult, we propose here another indirect measurement other than detecting
$\langle \sigma^{}_z \rangle$. The current is strictly located on the 2D
surface, so no in-plane magnetic field is induced by the current. On the other
hand, the spin vortices induce the in-plane magnetic moment due to the
existence of the SOCs as shown in Fig.~\ref{fig7} and Fig.~\ref{fig8}.

When both of the SOCs are present, examples of the magnetic induction lines
distributions are also indicated in Fig.~\ref{fig8}. In a small external
magnetic field, $\left\vert \mathbf{B}^{c}\right\vert \propto \left| g_1^2
-g_2^2\right|$. But it can be increased significantly with the increase of
the external magnetic field. According to Eqs.~(\ref{jx}), (\ref{jy}) and
(\ref{bslaw}), the current varies linearly with the external magnetic field,
so that the current induced magnetic is sensitive to the external magnetic
field. Consequently, from $B=0.1$T to $B=8$T, $|\mathbf{B}^c|$ increases by two
orders of magnitude. In contrast, the spin magnetic moment
does not vary much $\left\vert \mathbf{B}^{s}\right\vert \propto
\frac1{\hbar\omega r_0^2}$, since the magnitude of the spin field
in Eqs.~(\ref{sigmax}) and (\ref{sigmay}) is not strongly related to the
magnetic field. Therefore, in a strong magnetic field, the total induced
magnetic field above the center of the ring points down (Figs.~\ref{fig8}(b)
and (c)) due to the large cyclotron motion of the charge.

These findings in this subsection help us to find another way to measure the
topological charge of the in-plane field indirectly in a weak magnetic field,
i.e., to determine the direction of the total induced magnetic field above
the center of the ring as shown in Fig.~\ref{fig7}.

\section{Size effect and the limitations of the 1D model}

\subsection{The spin fields in different size of the ring}

The full Hamiltonian is difficult to solve analytically. However, in some
cases, we can have analytical approaches that are very close to the real
situation. The 2D Hamiltonian can be written as
\begin{equation}
H=\frac1{2m^*}\left[ \mathbf{p}+e\left(\mathbf{A+A}^R+\mathbf{A}^D\right)
\right]^2 + V(\mathbf{r}),
\end{equation}
where $V(\mathbf{r})$ is the confinement potential and the SOCs are written
in the form of effective vector potentials
$\mathbf{A}^R=\frac{mg^{}_1}{e}\left(-\sigma ^{}_y,\sigma ^{}_x \right),
\mathbf{A}^D=\frac{mg^{}_2}{e}\left(-\sigma ^{}_x,\sigma ^{}_y \right),$
for the Rashba and Dresselhaus SOCs, respectively. When the SOCs are weak,
the wave function is given approximately by
\begin{equation} \label{approwf}
\psi \left( \mathbf{r} \right) \approx \exp \left[ -i\frac{e}{\hbar } \left(
\mathbf{A}^R + \mathbf{A}^D \right) \mathbf{\cdot r}\right] \psi ^{}_0
\left( \mathbf{r} \right),
\end{equation}
where $\psi^{}_0$ is the wave function of the Hamiltonian without the SOC. In an
InAs quantum dot, $\psi^{}_0=\left(
               \begin{array}{cc}
                 \psi^{}_{n,l} & 0 \\
               \end{array}
             \right)^T$
is the wave function of the Fock-Darwin basis given by Eq.~(\ref{fd}).
We can obtain that the results of the special case that $g^{}_1=g^{}_2$ is
exactly the same as found in Ref.~\cite{luo1}.

For simplicity and without loss of generality, we choose $\psi^{}_0\left(
\mathbf{r}\right) =\left(
\begin{array}{cc}
f(\mathbf{r}) & 0
\end{array}
\right)^{T}$ where $f$ is the wave function of the quantum ring without the SOC.
We can then readily obtain
\begin{equation}
\psi \left( \mathbf{r} \right) =f(\mathbf{r}) \left(
\begin{array}{c}
\cos \Theta \\
\frac{-\beta^{}_1 e^{i\theta }+i\beta^{}_2e^{-i\theta }}{\Theta}
\sin \Theta
\end{array}%
\right) \allowbreak,
\end{equation}
where $\beta^{}_{1,2}=g^{}_{1,2}m^{\ast }r /\hbar$ should be smaller than
unity, and $\Theta=\sqrt{\beta_1^2-2\left( \sin 2\theta \right) \beta ^{}_1
\beta^{}_2+\beta_2^2} $. The spin fields are consequently
\begin{eqnarray}
\sigma^{}_x\left(r, \theta\right)  &=&- |f|^2 \frac{\sin 2\Theta}{\Theta}
 \left(\beta^{}_1\cos \theta -\beta^{}_2\sin \theta \right), \\
\sigma^{}_y\left(r, \theta \right) &=& |f|^2\frac{\sin 2\Theta}{\Theta}
 \left( \beta^{}_2\cos \theta -\beta^{}_1\sin \theta \right), \\
\sigma ^{}_z\left(r, \theta \right) &=& |f|^2 \cos 2 \Theta.
\end{eqnarray}

If the system without SOC has rotational symmetry, i.e. $|f|$ is independent
of $\theta$, the spin field can be simplified. If only the Rashba SOC is
present, then $\beta^{}_1\neq 0,\beta^{}_2=0,$
we have $\mathbf{\sigma }\left( \theta \right) \propto \left( -\sin 2\beta ^{}_1\cos
\theta,-\sin 2\beta^{}_1\sin \theta ,\cos 2\beta^{}_1\right) $ with
topological charge $q=1$, which is compatible with the perturbation
calculations. If $g^{}_1$ or $r^{}_0$ is small, then $\sin 2\beta^{}_1,
\cos 2\beta^{}_1 > 0$,
we can find that the in-plane spin vector $\left( -\sin 2\beta ^{}_1\cos
\theta,-\sin 2\beta ^{}_1\sin \theta \right)$
all points to the center and $\sigma^{}_z(\theta) >0$. When $\beta^{}_1 >
\pi/4$, $\cos 2\beta ^{}_1$ could be negative,
then $\sigma^{}_z(\theta) $ rotate to be negative. If $g^{}_1$ or $r^{}_0$
continues to increase then the spin vector rotates along the tangential line
of the ring.  However, if $g^{}_1$ or $r^{}_0$ increases too much, the wave function
obtained in Eq.~(\ref{approwf}) is no longer correct. Only the numerical
results in the next section are reliable.

In the similar manner, if only the Dresselhaus SOC is present, then we find
$\mathbf{\sigma }^{D} \propto\left( \sin 2\beta^{}_2 \sin \theta ,\sin 2\beta^{}_2
\cos \theta, \cos 2\beta^{}_2 \right)$ with $q=-1$. The results are compatible
with the quantum dot cases. If $g^{}_2$ or $r^{}_0$ increases, then the spin
vector also rotates. The analytical solution of the 1D ring refers to the SOCs
can be found in Ref.~\cite{sheng,arxiv}. However, no rotation of
$\sigma^{}_z(r)$ was found in the 1D model there. We shall see from our
numerical works that the rotation of $\sigma^{}_z(r)$ in a 2D ring is possible
for the Rashba and the Dresselhaus cases. The numerical results are beyond
the perturbation calculations where the SOCs and the radius ($\beta^{}_{1,2}$)
are treated as the small quantities, and are always reliable.

\subsection{Numerical results}

\begin{figure}[ptb]%
\centering
\includegraphics[
height=4.0542in,
width=2.8037in
]%
{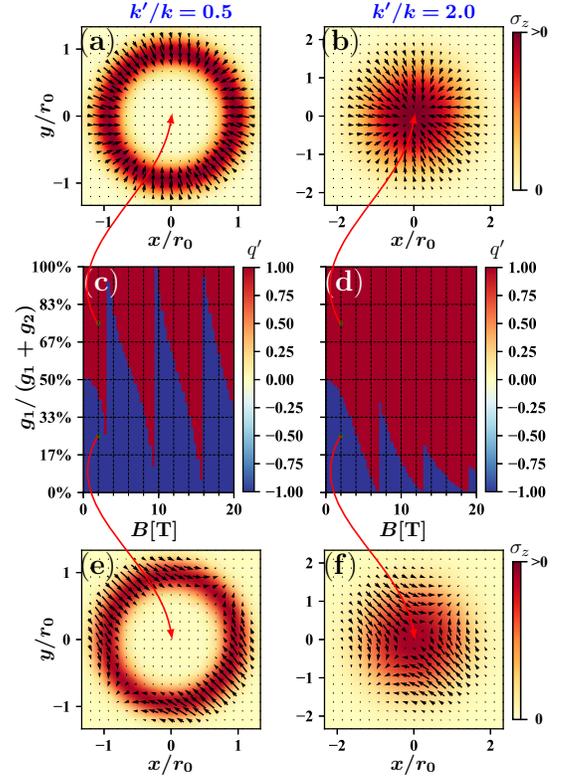}%
\caption{(Color online) The spin textures, density profiles and the
topological charges of a narrow ring [(a), (c), (e)], and a wide ring
[(b), (d), (f)], respectively. }
\label{fig9}%
\end{figure}

As discussed above, the spin rotates when $\beta^{}_{1,2}$ increases.
There are two possible ways to tune $\beta^{}_{1,2}$: Firstly, we can increase
the strength of the Rashba SOC which can be achieved by increasing the
potential of the gate \cite{Rash01,Rash03,Rash04,Rash05}. Secondly, the radius
can also be increased. In this subsection, We would like to explore how the
width and the radius of the ring affects the spin textures of the quantum ring.

For convenience, we define the ratio of the ring width $\sqrt{\frac{\hbar}
{m^* \omega}} $ to the radius $r^{}_0$, $k' (\hbar \omega,r^{}_0)=\sqrt{\frac
{\hbar} {m^{\ast} \omega r_{0}^2}}$. If $k'$ is small, the ring is
effectively narrow and the radius is large, while the ring is like a dot if
$k'$ is large. We choose the example $r^{}_0=15$ nm, $\hbar \omega=40$ meV in an
InAs ring as the reference, of which the ratio is $k\equiv k'(40,15)=0.449$.
We then explore how the spin textures evolves by changing this ratio.

For the ratio $k'=k$, the topological charges varying in the external
magnetic fields has been studied in Fig.~\ref{fig2}.
In Figs.~\ref{fig9}(a), (c), (e), spin textures, density profiles and the
topological charge $q'$ are displayed when the width-radius ratio is $k'=k/2$,
which corresponds to a big narrow ring. In Figs. \ref{fig9}(b), (d), (f),
those quantities are displayed when the ratio is $k'=2k$, corresponding to a
wide ring which is similar to a quantum dot. The periodic topological effect
due to the Aharonov-Bohm effect is still existing. However, in a wide ring,
there is no topological transition when $g^{}_1 > g^{}_2$, which is similar to the
case of quantum dot.

\begin{figure}[ptb]
\centering \includegraphics[width=3.2041in,height=3.6919in]{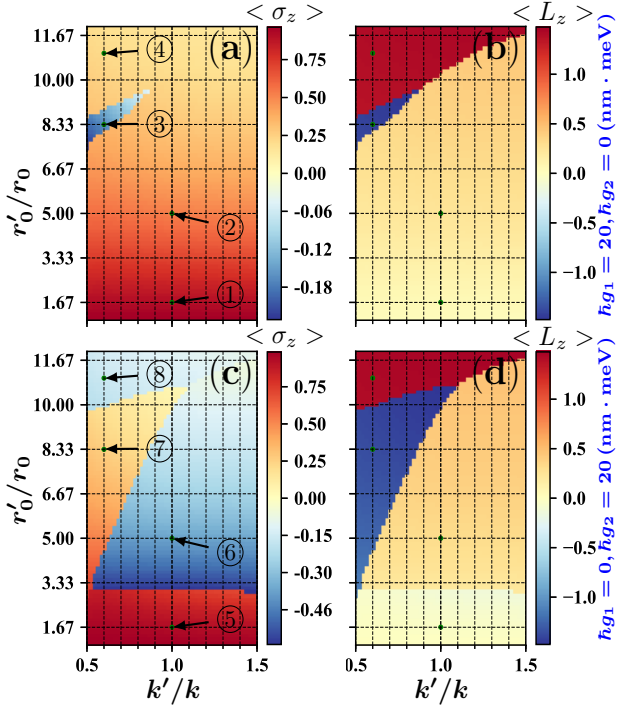}
\caption{(color online). The $\left\langle \sigma^{}_z\right\rangle $ and $\left\langle
L^{}_z\right\rangle $ of the ground state of InAs rings with $k^{\prime}/k\in
${[}0.5, 1.5{]}, $r_0^{\prime}/r_0^{}\in${[}0.5, 12.0{]}. The SOCs are:
(a) and (b) $ \hbar g^{}_1$ = 20 nm$\cdot$ meV, (c) and (d) $ \hbar g^{}_2$ =
20 nm$\cdot$ meV.}
\label{fig10}%
\end{figure}

We further study how the ratio affects the observable quantities such as
$\langle \sigma^{}_z \rangle$ and $\langle L^{}_z \rangle$. We utilize the
1D model in the perturbation theory. We find that the perturbative ground
state may vary when the radius is increased for a small magnetic field. The
energy of the unperturbed ground state without SOC is $E^{}_{0,+}<E^{}_{0,-}$
given by Eq. (\ref{ringenergy}). However, with the SOC, the real ground state
would be spin flipped, as in a quantum dot \cite{luo1}. Generally, the flip
condition is given by $E^{}_{0,+}+E_{0,+}^{(2)} > E^{}_{0,-}+E_{0,-}^{(2)}$.
In this case, the ground state will be altered to the spin-down state and
the $\langle \sigma^{}_z \rangle$ is flipped, which requires that the radius
satisfies ($g<0$)
\begin{equation}
r^{}_0>\frac{\hbar }{m^{\ast }}\sqrt{\frac{-0.5\beta ^{}_3}{g_2^2\left(
1-\beta ^{}_3\right) -g_1^2\left( 1+\beta^{}_3\right) }}
\end{equation}
when the magnetic field approaches zero, where $\beta^{}_3 = g m^*$.

If there is only the Dresselhaus SOC present, then we find that $r^{}_0>
\frac{\hbar}{g^{}_2 m^*} \sqrt{\frac{-\beta^{}_3} {2(1-\beta^{}_3)} }$. We
use the condition in Fig.~\ref{fig10} where we fixed the quantum flux and
$N=\frac{2eBr^{}_0}{\hbar}=0.34$, it results in $r^{}_0>39$ nm. This
agrees with the lowest transition line in Fig.~\ref{fig10}(c) where the
transition line of $\langle \sigma^{}_z \rangle$ is $r'_0 \approx 45$ nm. We
note that for the ring with Rashba SOC only, such a spin flipping also exists
when the radius of the ring increases, but for $\beta^{}_3 <-1$. We note that
for InAs, $\beta^{}_3 =-0.588 > -1$, so that this spin
flipping can not be found in the Rashba spin-orbit coupled 1D InAs ring. But
it is possible for some another materials with $g<0$, which is different from
the case of quantum dot. In quantum dots, the spin flipping is only possible
by the Dresselhaus SOC for a negative Land\'e $g$ material, and the Rashba SOC
can not flip the spin unless $g>0$ \cite{luo1}.

Surprisingly, we also numerically find the spin flipping happening in a
Rashba spin-orbit coupled 2D InAs ring with $g^{}_1 \neq 0, g^{}_2=0$ in Fig.
\ref{fig10}(a). It is completely different from the case of the InAs
quantum dot \cite{luo1} where only the Dresselhaus SOC can flip $\langle
\sigma^{}_z \rangle$ in a large dot. Moreover, we find another spin flipping
region in the Dresselhaus spin-orbit coupled 2D ring, as the triangle
region shown in Fig. \ref{fig10}(c). Those spin flips can not be
explained by the 1D model, where the radial differential is a constant
$\partial /\partial r \rightarrow 1/r_0$. However, in a 2D system this
term is never a constant. Only the 2D model stated in the last subsection
can qualitatively explain the rotation, i.e. $\cos 2\beta^{}_{1,2} <0$.
However, these abnormal spin flip regions do not correspond exactly to
$\beta^{}_{1,2} < \pi/4$, since the wave function obtained in
Eq.~(\ref{approwf}) is not accurate when the radius is large.

\begin{figure*}[ptb]%
\centering
\includegraphics
[
scale=0.9
]
{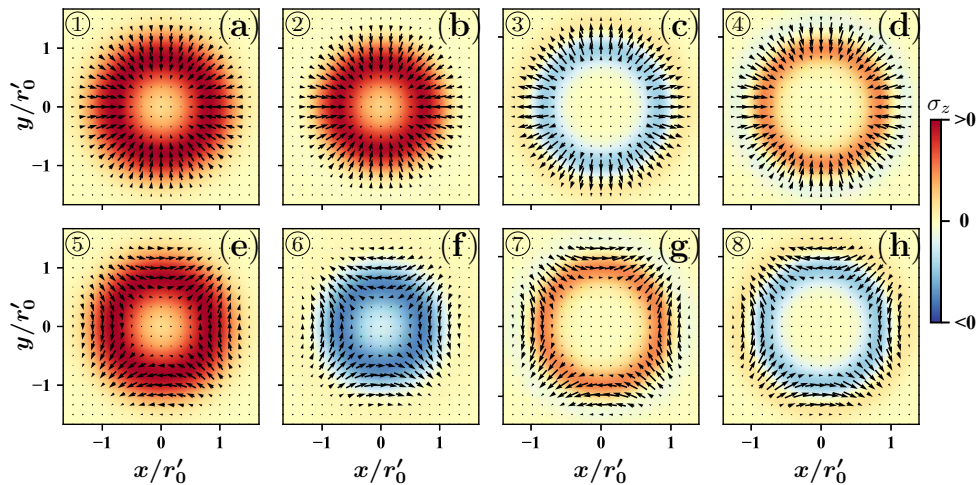}%
\caption{(color online) Spin textures of the ground states of InAs
rings (\textcircled{1} to \textcircled{8}) marked in Fig. \ref{fig10}.}
\label{fig11}
\end{figure*}

In Fig. \ref{fig10}, the flipping of $\langle \sigma^{}_z \rangle$ (and 
$\langle L^{}_z \rangle$) in the rings with Rashba SOC or with Dresselhaus 
SOC are shown, which may be observed in NMR experiments. The spin textures of 
some cases marked in Fig. \ref{fig10}, \textcircled{1} to \textcircled{8} are
plotted in Fig. \ref{fig11}. In the case \textcircled{6} shown in Fig. 
\ref{fig11}(f) where the spin 
flipping is induced by the energy alternating $E^{}_{0,+}+E_{0,+}^{(2)} > 
E^{}_{0,-}+E_{0,-}^{(2)}$, so that all the $\sigma_z(\mathbf{r})$ points 
down. However in Figs. \ref{fig11}(c), (d), (g) and (h), with the increase 
of the radius we can clearly observe that $\langle \sigma^{}_z \rangle$ 
changes its sign by the size effect, which can not be explained by the 1D
model.

\section{Conclusion}

In summary, we have studied the 1D ideal model if the SOCs are weak and
can be treated as perturbation. We analyze the system in the perturbation
theory. We then show that the winding number (topological charge) of the
InAs quantum rings can be tuned by both electric and magnetic fields in
the presence of both Rashba and Dresselhaus SOCs. Compared to the quantum dots,
quantum rings exhibit several new properties, the major one is that that its
winding number periodically changes the sign with the magnetic field due
to the AB effect.

The current induced by magnetic field and by the SOCs reveal nonzero
persistent current $I^{}_{total}$. $I^{}_{SOC}$ plays an important role which
makes it possible to distinguish the type of SOC by adjusting the current
direction when $l-N\approx0$ in a weak magnetic field. Consequently, the
magnetic fields induced by these currents and the spin field are
also numerically calculated and we find that it may be used as a means of
measuring the topology of the spin textures in a weak external magnetic field.
Some of the spin rotations with the increase of the radius can be found only
in the 2D model, as the 1D model is inadequate for the large ring radius.
The size effect can be observed by changing the ring's radius $r^{}_0$ and the
width, if the strength of the SOC is difficult to tune. The direction of the
spin field could be changed with the increase of $r^{}_0$, when the width is
narrow relative to the radius. With only the Rashba SOC ithe spin direction
cannot be changed in InAs quantum dots, but it can be changed in quantum ring
by increasing $r^{}_0$. With Dresselhaus SOC only, spin direction changes more
than once in the quantum rings when the radius is increased.
These findings pave the way to control the topological features of the system
in spintronics \cite{Zutic,Smejkal} and may be useful in quantum computation
\cite{qbit}. It also leads to the findings of the transport properties when
the quantum ring is designed as a transport device \cite{peng}.

\section{Acknowledgement}

This work has been supported by the NSF-China under Grant No. 11804396.
F.O. acknowledges financial support by the NSF-China under Grant No. 51272291,
the Distinguished Young Scholar Foundation of Hunan Province (Grant No. 2015JJ1020),
and the CSU Research Fund for Sheng-hua scholars (Grant No. 502033019).

\end{document}